\documentclass[preprint,nofootinbib,12pt,floatfix,superscriptaddress,aps,prd]{revtex4-2}
\usepackage{amsmath}
\usepackage{amssymb}
\usepackage{amsbsy}
\usepackage{amsfonts}
\usepackage{amsopn}
\usepackage{graphicx}
\usepackage{amssymb}
\usepackage{amsfonts}
\usepackage{amsmath}
\usepackage{graphicx}
\usepackage{color}
\usepackage{slashed}
\usepackage{esint}
\usepackage[dvips]{epsfig}
\usepackage[dvips]{graphicx}
\usepackage{float}
\usepackage{units}
\usepackage{textcomp}
\usepackage{hyperref}
\usepackage[caption=false]{subfig}
\newcommand{\beq}{\begin{equation}}
\newcommand{\eeq}{\end{equation}}
\newcommand{\bea}{\begin{eqnarray}}
\newcommand{\eea}{\end{eqnarray}}

\begin{document}

\title{Lower-dimensional Gauss–Bonnet gravity black holes with quintessence}

\author{G. Alencar}
\email{geova@fisica.ufc.br}
\affiliation{Departamento de F\'isica, Universidade Federal do Cear\'a, Caixa Postal 6030, Campus do Pici, 60455-760 Fortaleza, Cear\'a, Brazil.}

\author{T. M. Crispim}
\email{tiago.crispim@fisica.ufc.br}
\affiliation{Departamento de F\'isica, Universidade Federal do Cear\'a, Caixa Postal 6030, Campus do Pici, 60455-760 Fortaleza, Cear\'a, Brazil.}

\author{J. Macedo}
\email{joaomacedocabral@fisica.ufc.br}
\affiliation{Departamento de F\'isica, Universidade Federal do Cear\'a, Caixa Postal 6030, Campus do Pici, 60455-760 Fortaleza, Cear\'a, Brazil.}

\author{C. R. Muniz}
\email{celio.muniz@uece.br}
\affiliation{Universidade Estadual do Cear\'a, Faculdade de Educa\c c\~ao, Ci\^encias e Letras de Iguatu, 63500-000, Iguatu, CE, Brazil.}

\begin{abstract}
In this paper, we study the $D\to3$ limit of Gauss–Bonnet gravity with quintessential matter, obtaining exact solutions that extend the BTZ metric through higher-curvature terms and quintessence coupling. The solutions exhibit a single event horizon whose radius decreases with increasing quintessence parameter $\omega_q$, while developing a curvature singularity at the origin for non-vanishing quintessence. The geodesic analysis reveals stable circular photon orbits exist exclusively for phantom-like quintessence ($\omega_q < -1$). Thermodynamically, the system is stable, since the specific heat is positive, and with evaporation it evolves to stable remnants whose characteristic size decreases as $\omega_q$ increases, with complete evaporation prevented by quintessence effects. Furthermore, we find that all physical quantities intrinsically depend on the parameter $\alpha$ of the Gauss-Bonnet extension.These results demonstrate the profound influence of quintessential matter on both geometric and thermodynamic properties of (2+1)-dimensional black holes, offering new perspectives on gravitational theories in lower dimensions and black hole final states. 
\end{abstract}

\maketitle

\newpage

\section{Introduction}
Lovelock gravity represents the most general extension of General Relativity (GR) constructed solely from the metric and the Riemann curvature tensor, while preserving second-order field equations for the metric~\cite{Lovelock:1971yv}. The theory generalizes Einstein's framework by including, in addition to the cosmological constant and the Einstein--Hilbert action, a series of higher-order curvature invariants that contribute nontrivially only in spacetimes with $D>4$. The first of these invariants is the Gauss--Bonnet (GB) term, given by
\begin{equation}
\mathcal{G} = R_{abcd}R^{abcd} - 4 R_{ab}R^{ab} + R^2\,,
\label{GBinv}
\end{equation}
which affects the dynamics only in a spacetime with $D \geq 5$. For lower dimensional spacetimes, these higher-order terms are either topological or identically vanish. As a result, in four dimensions, GR remains the unique metric theory of gravity yielding second-order field equations.

A recent proposal~\cite{Glavan:2019inb} has sparked significant interest in the possibility of overcoming the constraints imposed by Lovelock’s theorem. The core idea involves treating the spacetime dimension $D$ as a continuous parameter and performing a dimensional regularization procedure. Specifically, one rescales the GB coupling as
\begin{equation}
(D - 4)\alpha \to \alpha\,,
\end{equation}
and then takes the formal limit $D \to 4$ to obtain a nontrivial contribution from the GB term in four dimensions. As originally formulated, this approach constructs four-dimensional solutions as limiting cases of higher-dimensional ones, thereby inheriting corrections from higher-curvature terms. This method has led to a wide range of $D = 4$ black holes, wormholes and cosmological solutions that retain imprints of higher-dimensional geometry, and has been extensively explored in the literature~\cite{Glavan:2019inb,Kumar:2020uyz,Fernandes:2020rpa,Kumar:2020owy,Kumar:2020xvu,Li:2020tlo,Kobayashi:2020wqy,Doneva:2020ped,Ghosh:2020vpc,Malafarina:2020pvl,Casalino:2020kbt,Konoplya:2020qqh,Konoplya:2020der,EslamPanah:2020hoj,Konoplya:2020cbv,Wei:2020poh,HosseiniMansoori:2020yfj,Zhang:2020qam,Singh:2020xju,Hegde:2020cdm,Konoplya:2020bxa,Zhang:2020qew,Zhang:2020sjh,Wei:2020ght,Guo:2020zmf,Jin:2020emq,Heydari-Fard:2020sib,Liu:2020vkh,Islam:2020xmy,Roy:2020dyy,NaveenaKumara:2020kpz,Mishra:2020gce,Yang:2020czk,Ge:2020tid,Carvalho:2022ywl,Cunha:2025jzh}.

Following studies have criticized the consistency and interpretation of this approach~\cite{Gurses:2020ofy, Shu:2020cjw, Hennigar:2020lsl, Tian:2020nzb, Mahapatra:2020rds, Bonifacio:2020vbk, Arrechea:2020evj}\cite{Fernandes:2022zrq}. For instance, Ref.~\cite{Gurses:2020ofy} demonstrated that no well-defined geometric object can be identified to serve as the field equations of the resulting four-dimensional theory. Meanwhile, Refs.~\cite{Hennigar:2020lsl, Tian:2020nzb} examined more intricate settings, including cosmological and Taub--NUT spacetimes, and found that the limiting procedure $D \to 4$ is inherently non-unique. Specifically, there exist multiple inequivalent ways to embed a four-dimensional metric into higher dimensions, and the limiting solutions may retain residual dependencies on the geometry of the extra dimensions.

These challenges can be addressed by carefully taking the limits of the higher-dimensional theory itself. L{\"u} and Pang~\cite{Lu:2020iav} (and similarly, \cite{Kobayashi:2020wqy}) employed a Kaluza-Klein-like procedure to derive a four-dimensional limit of GB gravity by compactifying the higher-dimensional theory on a maximally symmetric space, followed by taking the limit where the dimension of the space vanishes. Later, an alternative approach was proposed by~\cite{Fernandes:2020nbq, Hennigar:2020lsl}, which is a generalization of a technique introduced nearly three decades ago by Mann and Ross to derive a $D \to 2$ limit of general relativity~\cite{Mann:1992ar}. This method aims to derive a $D \to 4$ limit of GB gravity without making any assumptions about the nature of the extra dimensions. Interestingly, both approaches lead to the same limiting theory~\cite{Hennigar:2020fkv}.

The action for this theory is given by
\bea\label{SD}
S&=&\int d^D x \sqrt{-g}\Bigl[R-2\Lambda+\alpha\Bigl(\phi {\cal G}+4 G^{ab}\partial_a \phi \partial_b \phi\nonumber\\
&&\qquad\qquad\qquad -4(\partial \phi)^2 \Box \phi+2((\nabla\phi)^2)^2 \Bigr)+\mathcal{L}_M \Bigr]\,,
\eea
where $\mathcal{L}_M$ represents the matter fields, particularly the quintessential fluid considered in this context. This action~\eqref{SD} can be interpreted as the closest equivalent of GB gravity in four dimensions. The theory includes an additional scalar degree of freedom and is a special case of Horndeski theory~\cite{Horndeski:2024hee}, with \cite{Schmidt:2018zmb} providing a construction of conserved currents in these theories. The naive $D \to 4$ limit of the higher-dimensional spherically symmetric black hole solution in GB gravity also applies to this theory for a specific scalar configuration, although it does not represent the most general solution. The structure of more complex solutions, such as Taub-NUT, is more intricate and does not correspond directly to the limits of the higher-dimensional solutions~\cite{Hennigar:2020lsl}.

While considerable attention has been focused on understanding the four-dimensional limit of GB gravity, the BTZ black hole \cite{Banados:1992wn} has provided a fertile ground for solving many problems that are intractable in higher dimensions~\cite{Ross:1992ba, Carlip:1995qv, Cardoso:2001hn, Konoplya:2004ik, Townsend:2013ela, Shu:2014eza, Bravo-Gaete:2014haa, Chernicoff:2018hpb}. In this regard, the authors of Ref. \cite{Hennigar:2020fkv} applied it to 3D and derived a generalization of the BTZ solution, given by
\beq \label{NewSolution}
f(r)=-\frac{r^2}{2\alpha}\left[1- \sqrt{1+\frac{4\alpha}{r^2}\left(\frac{r^2}{\ell^2}-m\right)} \right],
\eeq 
which recovers the usual solution in the limit $\alpha \to 0$ \cite{Banados:1992wn}.

However, to the best of our knowledge, the 3D case with sources has yet to be explored in the literature. Motivated by this gap, we explore the action (\ref{SD}) in three dimensions, considering the presence of an anisotropic fluid of quintessence type. Black holes, neutral and charged, sourced by quintessence are studied in \cite{Kiselev:2002dx, Chen:2012mva, Cui:2021wxn}. Furthermore, various studies have explored the interplay between GB terms and quintessence-like fluids. In four-dimensional contexts, black hole solutions with these features have been discussed in~\cite{Shah:2021rob,Ladghami:2023kyb}, where modifications due to the pressure of dark energy components were found to impact horizon structure and thermodynamic stability. Charged BTZ black holes are discussed in \cite{Jusufi:2023fpo}. In particular, \cite{Ahmed:2022dpu} showed that the weak cosmic censorship conjecture is not valid in this context. 

Thus, motivated by previous studies~\cite{Hennigar:2020fkv}, which treated certain physical quantities as independent of the Einstein-Gauss-Bonnet(EGB) parameter $\alpha$, our original goal was to generalize these results to the case with a source (quintessence). Upon revisiting the problem, we found that, contrary to the prior assumption, all physical quantities inherently depend on $\alpha$, even in the unsourced case. This observation, naturally extends to the case with quintessence, establishing the $\alpha$-dependence in the generalized scenario. 

The manuscript is organized as follows: In Section II, we derive the BTZ black hole solution surrounded by a quintessence fluid in GB gravity and analyze its horizon structure. In Section III, we examine the geodesic circular orbits within this spacetime. In Section IV, we explore the thermodynamics of the black hole. Finally, in Section V, we summarize the paper and conclude it.

\section{BTZ Sourced by quintessence}
In this section, we revisit the solution of Ref.~\cite{Hennigar:2020fkv}, where it was argued that the relevant quantities do not depend on the parameter $\alpha$. In contrast, we argue that all physical quantities should indeed depend on $\alpha$. After these considerations, we proceed to analyze the case sourced by quintessence.

\subsection{Revisiting the Gauss–Bonnet BTZ black hole}

The authors of Ref.~\cite{Hennigar:2020fkv} obtained the GB generalization of the BTZ solution, which takes the form  
\beq \label{NewSolution}
f(r)=-\frac{r^2}{2\alpha}\left[1- \sqrt{1+\frac{4\alpha}{r^2}\left(\frac{r^2}{\ell^2}-C_1\right)} \,\right],
\eeq 
where $C_1$ is an integration constant. For large values of $r$, the asymptotic behavior of this metric was found to be~\cite{Hennigar:2020fkv}  
\beq 
f(r)\approx -\frac{C_1}{\sqrt{1+\frac{4\alpha}{\ell^2}}}+\frac{r^2}{2\alpha}\left(\sqrt{1+\frac{4\alpha}{\ell^2}}-1\right)=-M_{\text{eff}}+r^2\Lambda_{\text{eff}},
\eeq 
with the effective parameters defined as  
\begin{equation}
M_{\text{eff}}=\frac{C_1}{K}, \qquad \Lambda_{\text{eff}}=\frac{K-1}{2\alpha}, \qquad K=\sqrt{1+\frac{4\alpha}{\ell^2}}.
\end{equation}
This asymptotic form will be essential in the following discussion, where we argue that all physical quantities must in fact depend on the parameter $\alpha$.

To understand the meaning of the integration constant $C_1$, we consider the limit $\alpha \to 0$. In this case, one finds $C_1 = m_{BTZ}$, where $m_{BTZ}$ is the mass parameter of the usual BTZ solution. Although $C_1 = m_{BTZ}$ in this limit, the physically meaningful constant in the GB generalization is $M_{\text{eff}}$. Therefore, all physical quantities should be expressed in terms of $M_{\text{eff}}$. For instance, as described in Ref.~\cite{Hennigar:2020fkv}, the horizon radius was written as $r_+^2 = C_1 \ell^2 = m_{BTZ} \ell^2$, and the authors argued that the horizon radius coincides with that of the usual BTZ black hole. However, this interpretation is misleading, since the radius must be expressed in terms of the physical mass. Accordingly, we obtain
\beq
r_+^2 =   K M_{\text{eff}} \ell^2,
\eeq
which shows that the horizon radius depends explicitly on the GB parameter $\alpha$.  

Even though $M_{\text{eff}}$ is the relevant quantity, its relation with $m_{BTZ}$ remains significant, since the usual thermodynamics of the BTZ black hole is expected to remain valid. With this in mind, the modified first law of thermodynamics must satisfy 
\begin{equation}
   dM_{\text{eff}} = T_{\text{eff}}\, dS_{\text{eff}},
\end{equation}
and, therefore,
\begin{equation}\label{modified}
   dm_{BTZ} = K\, T_{\text{eff}} dS_{\text{eff}} = T_{BTZ}\, dS_{BTZ} 
   \;\;\;\Rightarrow\;\;\;  
   T_{\text{eff}} dS_{\text{eff}} = \frac{1}{K}\, T_{BTZ} dS_{BTZ}.
\end{equation}
Therefore, as compared with the BTZ case, the thermodynamic quantities necessarily depend on $\alpha$. This result is in direct contradiction with the claim Ref.~\cite{Hennigar:2020fkv} that the thermodynamics of the system remains unchanged. Therefore, their conclusion must be regarded as a clear misinterpretation of the thermodynamic consistency conditions. This dependence is unavoidable once the effective mass $M_{\text{eff}}$ is identified as the true physical parameter of the solution.

However, by employing the Wald entropy, the authors of Ref.~\cite{Hennigar:2020fkv} concluded that the entropy remains unchanged. By using this and $T_{eff}=M'_{eff}/S'_{eff}$, we deduce that 
\begin{equation}
    T_{\text{eff}} = \frac{T_{BTZ}}{K}.
\end{equation}
This is consistent with Eq. (\ref{modified}).  This concludes our revisitation of the GB BTZ solution. The essential point is that the parameter $\alpha$ cannot be disregarded: it permeates not only the geometry but also the physical observables, including the horizon structure and thermodynamics. In the next section, we extend this analysis to the case sourced by quintessence.

\subsection{The case with source}
In this section, we will consider the action (\ref{SD}) and spherical symmetry, with metric 
\begin{equation}
    ds^2=-f(r)dt^2+\frac{1}{f(r)}dr^2+r^2d\varphi^2.
\end{equation}
The gravitational field equations are given by
\beq
\label{eisteineq}	G_{\mu}^{\;\nu}-\alpha \mathcal{H}_{\mu}^{\;\nu}=T_{\mu}^{\;\nu},
\eeq
where $G_{\mu\nu}$ is the Einstein tensor, $T_{\mu}^{\;\nu}$ is the Quintessence stress-energy tensor and $\mathcal{H}_{\mu\nu}$ is defined by
\bea 
\mathcal{H}_{\mu\nu}\!&\!=\!&\!2R\left(\nabla_\mu\nabla_\nu \phi-\nabla_\mu\phi\nabla_\nu\phi\right)+2G_{\mu\nu}\left[(\nabla\phi)^2-2\Box\phi\right]+4G_{\nu\alpha}\left(\nabla^\alpha\nabla_\mu\phi-\nabla^\alpha\phi\nabla_\mu\phi\right)\nonumber\\
&&+4G_{\mu\alpha}\left(\nabla^\alpha\nabla_\nu\phi-\nabla^\alpha\phi\nabla_\nu\phi\right)
+4R_{\mu\alpha\nu\beta}\left(\nabla^\beta\nabla^\alpha\phi-\nabla^\alpha\phi\nabla^\beta\phi\right)+4\nabla_\alpha\nabla_\nu\phi\left(\nabla^\alpha\phi\nabla_\mu\phi-\nabla^\alpha\nabla_\mu\phi\right)\nonumber\\
&&+4\nabla_\alpha\nabla_\mu\phi\nabla^\alpha\phi\nabla_\nu\phi+4\Box\phi\nabla_\nu\nabla_\mu\phi
-4\nabla_\mu\phi\nabla_\nu\phi\left[(\nabla\phi)^2+\Box\phi\right]^2-g_{\mu\nu}\Big\{2R\left[\Box\phi-(\nabla\phi)^2\right]\nonumber\\
&&+4G^{\alpha\beta}\left(\nabla_\beta\nabla_\alpha\phi-\nabla_\alpha\phi\nabla_\beta\phi\right)+2(\Box\phi)^2-(\nabla\phi)^4+2\nabla_\beta\nabla_\alpha\phi\left(2\nabla^\alpha\phi\nabla^\beta\phi-\nabla^\beta\nabla^\alpha\phi\right)\Big\}.
\eea
The field equation of the scalar field $\phi$ reads:
\bea \label{r4EGBgfeq1}
	&&R^{\mu\nu}\nabla_{\mu}\phi\nabla_{\nu}\phi-G^{\mu\nu}\nabla_{\mu}\nabla_{\nu}\phi-\Box\phi\nabla_{\mu}\phi\nabla^{\mu}\phi+\nonumber
    \\
    &&\nabla_{\mu}\nabla_{\nu}\phi\nabla^{\mu}\nabla^{\nu}\phi-(\Box\phi)^{2}-2\nabla_{\mu}\phi\nabla_{\nu}\phi\nabla^{\mu}\nabla^{\nu}\phi=0.
\eea

In coordinates, equations \eqref{eisteineq} and (\ref{r4EGBgfeq1}) are given by
\begin{align}
   & \Lambda + \frac{f'}{2r} + \frac{\alpha}{2r}\Big[2 f \phi'^2 \left(3 - 2 r \phi'\right) f' 
- 2 f^2 \phi' \left(r \phi'^3 - 4 \phi'' + 4 r \phi' \phi''\right)
\Big]=T_{t}^{\;t}&,
\\
& \Lambda + \frac{f'}{2r} + \frac{\alpha}{2r}\Big[2 f^2 \phi'^3 \left(-4 + 3 r \phi'\right) 
+ 2 f \phi'^2 \left(3 - 2 r \phi'\right) f'
\Big]=T_{r}^{\;r}&,
\\
&\Lambda + \frac{f''}{2} + \alpha \Big[ 
\phi'^2 f'^2 
- f^2 \phi'^2 \big(\phi'^2 + 4 \phi''\big) 
+ f \phi' \big(-2 \phi'^2 f' 
+ 2 f' \phi'' + \phi' f''\big)
\Big]=T_{\varphi}^{\;\varphi}&,\\
&\phi' ( -1 + r \phi' )f'^2 - 2 f^2 \phi' \left[  \phi'^2 + ( -2 + 3 r \phi' ) \phi'' \right]+&
\nonumber\\
&+ f \left[ - f' \left( \phi'' + \phi' \left( \phi' ( -5 + 4 r \phi' ) - 2 r \phi'' \right) \right)  \phi' ( -1 + r \phi' ) f'' \right] = 0&.
\end{align}
Now, we introduce the quintessence source, described by 
\begin{equation}\label{sources}
  T_{t}^{\;t}=T_{r}^{\;r}=  -\frac{\omega \kappa }{r^{2 (w+1)}}; T_{\varphi}^{\;\varphi}=-\frac{(2\omega+1)\omega \kappa  }{r^{2 (w+1)}}.
\end{equation}
By subtracting the $r-r$ and $t-t$ equations, we have
\begin{equation}
    \frac{4 \alpha f^2  \big( r \phi'^2 -\phi'\big) \big(\phi'^2 + \phi''\big)}{r}=0.
\end{equation}
The solution of the above equation is given by
\beq \label{profile}
\phi=\ln(r/l)\,,
\eeq 
where \( l \) is an integration constant, and its presence renders the spacetime no longer of constant curvature. With this, the equation from $\delta \phi$ is then identically satisfied. Finally, by summing the $r-r$ and $t-t$ equations we have
\beq 
\frac{\alpha f(r) f'(r)}{r^3}-\frac{\alpha f(r)^2}{r^4}+\frac{f'(r)}{2 r}-\frac{1}{\ell^2}=-\frac{ \kappa \omega }{r^{2 (w+1)}}.
\eeq 
The solutions are
\beq \label{NewSolution}
f_{\pm}(r)=-\frac{r^2}{2\alpha}\left[1\pm \sqrt{1+\frac{4\alpha}{r^2}\left(\frac{r^2}{\ell^2}-C_1-  q r^{-2\omega_q}\right)} \right].
\eeq 
where $C_1$ is an integration constant. We made $q=|\kappa \omega|$ in the above equation. Notice that only the $f_-$ branch recovers the solution (\ref{NewSolution}) without quintessence ($q=0$). This branch also has a limit 
\beq 
\lim_{\alpha \to 0}f(r)= -C_1+\frac{r^2}{\ell^2}-q r^{-2 \omega_q},
\eeq 
and $C_1=m_{BTZ}$ corresponds to the solution obtained in \cite{deOliveira:2018weu}. On the other hand, for $r\to\infty$, we have
\beq 
f(r)\approx \frac{r^2}{2\alpha}\left(K-1\right)-\frac{C_1}{K}-\frac{q r^{-2\omega_q}}{K},
\eeq 
and thus we must fix the constraint $-\ell^2/4<\alpha<0$ to have a well-defined black hole solution in all space. Outside this range of $\alpha$, $K$ becomes complex, making the model physically ill-defined. Therefore, we must have an adS-like solution with an effective cosmological constant and effesctive mass given by
\beq 
\Lambda_{eff}=\frac{K-1}{2\alpha},\,M_{eff} =\frac{m_{BTZ}}{K}.
\eeq 

We plot the behavior of $\Lambda_{eff}$ and $M_{eff}$ with respect to the GB parameter. We notice that both are monotonically decreasing and strictly positive.

\begin{figure}[h!]
    \centering
    \includegraphics[width=0.7\linewidth]{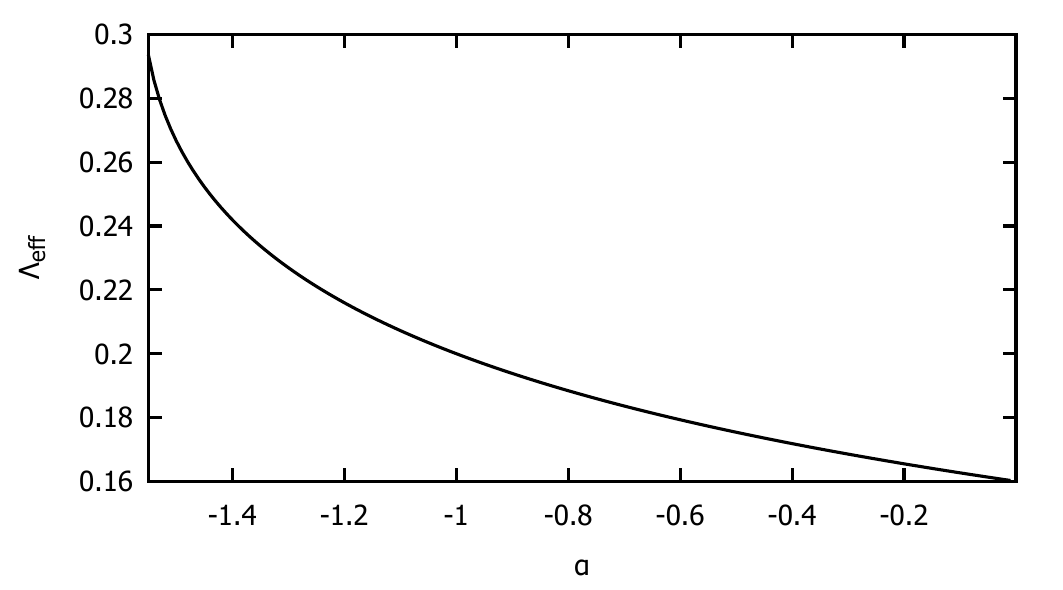}
    \includegraphics[width=0.7\linewidth]{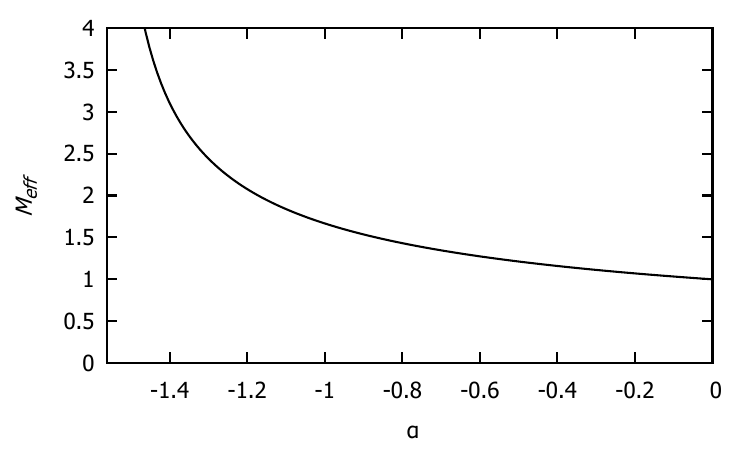}
    \caption{Effective cosmological constant (top panel) and effective mass (bottom panel) as a function of the GB parameter. The parameter setting is: $\ell=2.5$ and $m_{BTZ}=1$, in Planck units.}
    \label{fig:lambdaeff}
\end{figure}

The Ricci curvature, nearby the origin, $r=0$, is given by
\beq 
R\approx -\frac{6}{\ell^2}+\frac{6\alpha}{\ell^4}-2 q\omega_q\left(1-2\omega_q-\frac{\alpha}{\ell^2}+\frac{4\omega_q\alpha}{\ell^2}\right)r^{-2(\omega_q+1)}+4q\omega_q\alpha(1+3\omega_q+2\omega_q^2)r^{-2(\omega_q+2)},
\eeq 
for $q,\alpha\ll 1$. Notice that, as $\omega_q>-2$, the EGB (2+1) black holes surrounded by quintessence are not regular at the origin. If $q \to 0$, the regularity of the black hole is reestablished.

\begin{figure}[ht]
    \centering
    \includegraphics[width=0.7\linewidth]{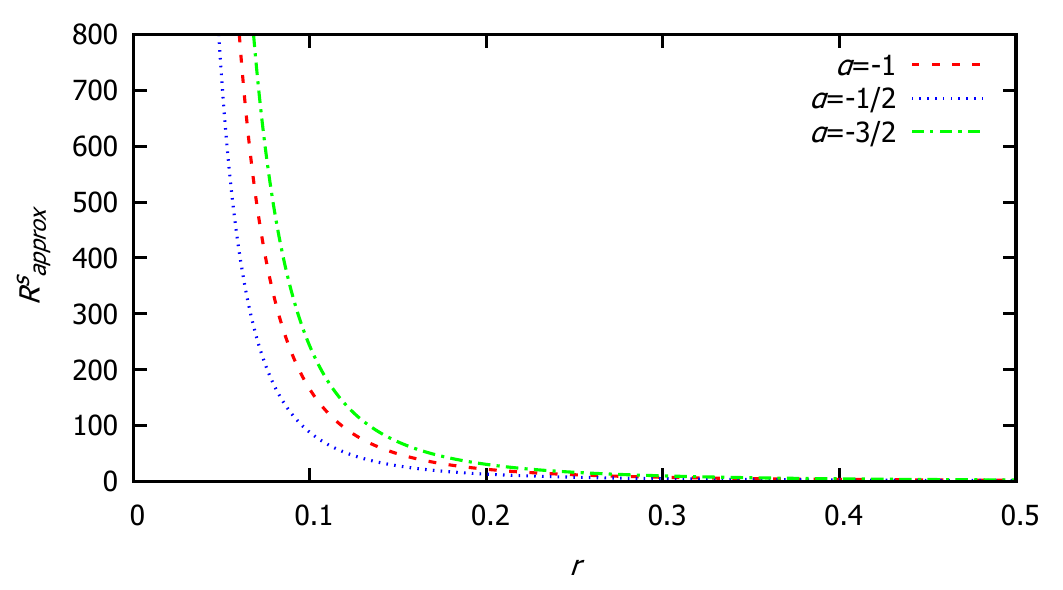}
     \includegraphics[width=0.7\linewidth]{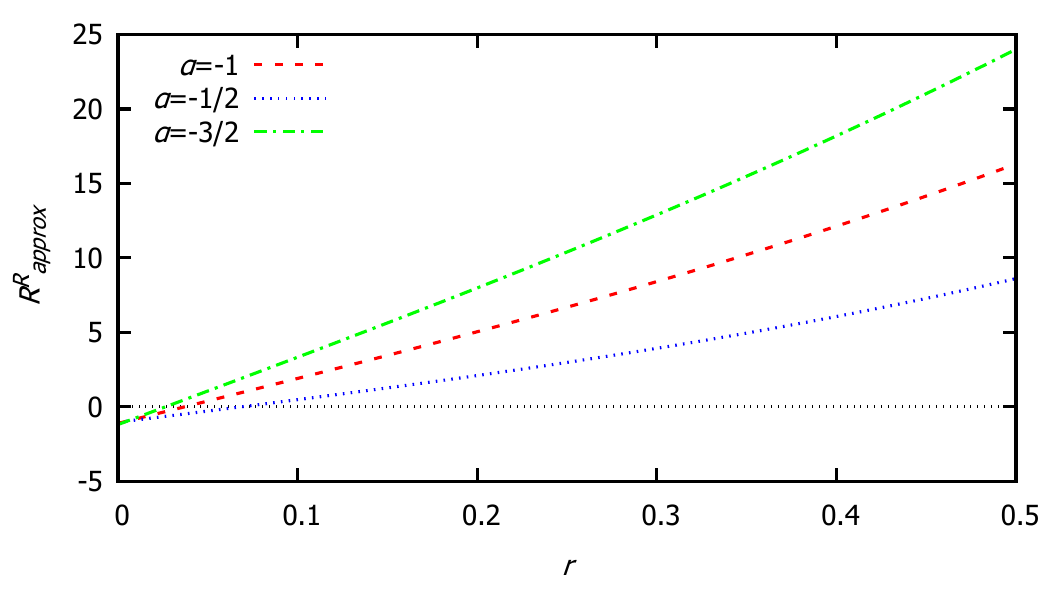}
    \caption{ Approximate Ricci scalar for $r$ near the origin as a function of the radial coordinate, for the singular case (top panel, $\omega = -2/5$) and regular case (bottom panel, $\omega = -5/2$) varying $\alpha$. The parameter settings are:$\ell=2.5$, in Planck units.}
    \label{fig:R}
\end{figure}

Furthermore, Fig. \ref{fig:R} shows the behavior of the approximate Ricci scalar for values of $r$ near the origin, for different values of $\alpha$, in the singular case (top panel) and regular case (bottom panel). From this, in the singular case, we see that the Ricci scalar converges to a finite value at $r\to\infty$ depending on $\alpha$. The role of the GB parameter, in this case, is to determine the asymptotic behavior of the Ricci scalar.

\begin{figure}[h!]
    \centering
            \includegraphics[width=0.7\textwidth]{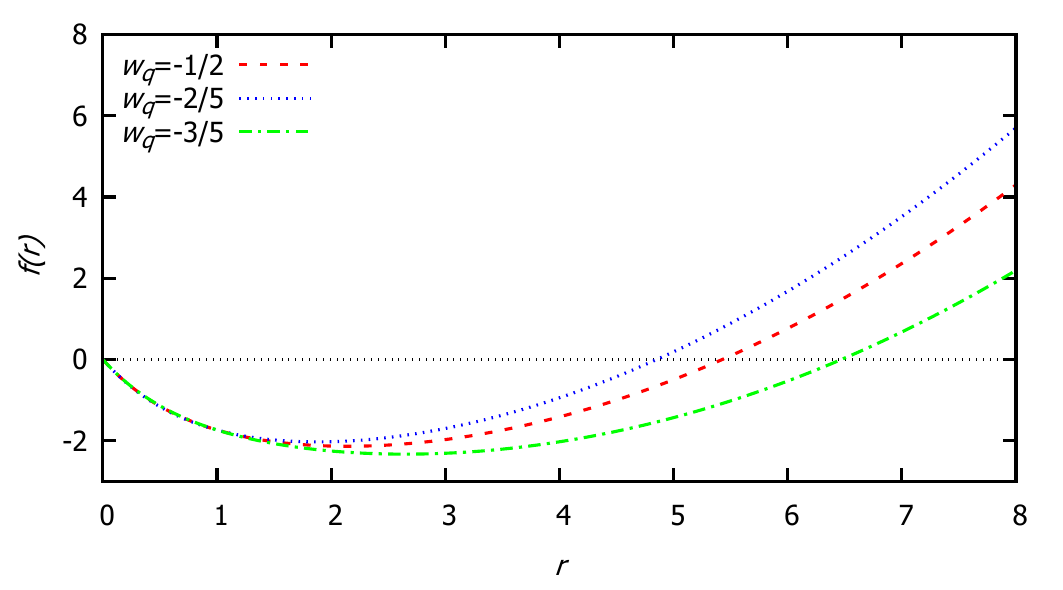}
        \caption{Metric coefficient as a function of the radial coordinate, $r$. The parameter settings are: $\alpha=-0.2$, $q=0.5$, $M=2.0$, and $\ell=2.5$, in Planck units, varying $\omega_q$}
    \label{MetricCoeff}
\end{figure}

Fig \ref{MetricCoeff} depicts the metric coefficient $f(r)$ as a function of the radial coordinate, where we can see the event horizon radii, $r_h$, at which $f(r)=0$. 

We can also see that the smaller the state parameter $\omega_q$, the greater the horizon radius. The horizon is defined by 
\beq 
\frac{r_h^2}{\ell^2}-K M_{eff}-q r_h^{-2\omega_q}=0,
\eeq
which, for $\omega_q=-1/2$, has the analytical solution 
\beq 
r_h=\frac{1}{2} \left(\ell^2 q+\sqrt{\ell^4 q^2+4 \ell^2 KM_{eff}}\right).
\eeq

\section{Geodesic circular orbits}

The geodesic of a particle in orbit around a static black hole is given by
\begin{equation}
\dot{r}^2=  \mathcal{E}^2 - f(r)\left(\frac{\lambda^2}{r^2}-\epsilon\right),
\end{equation}
where the dot over $r$ indicates the derivative with respect to the affine parameter $\mathcal{E}$ is the particle energy, $\lambda$ is the angular momentum, and $\epsilon=-1$ ($=0$) for a massive particle (for light).
The effective potential is defined as
\beq
V_r=f(r)\left(\frac{\lambda^2}{r^2}-\epsilon\right).
\eeq\label{effectpot}
\begin{figure}[htb!]
    \centering
            \includegraphics[width=0.7\textwidth]{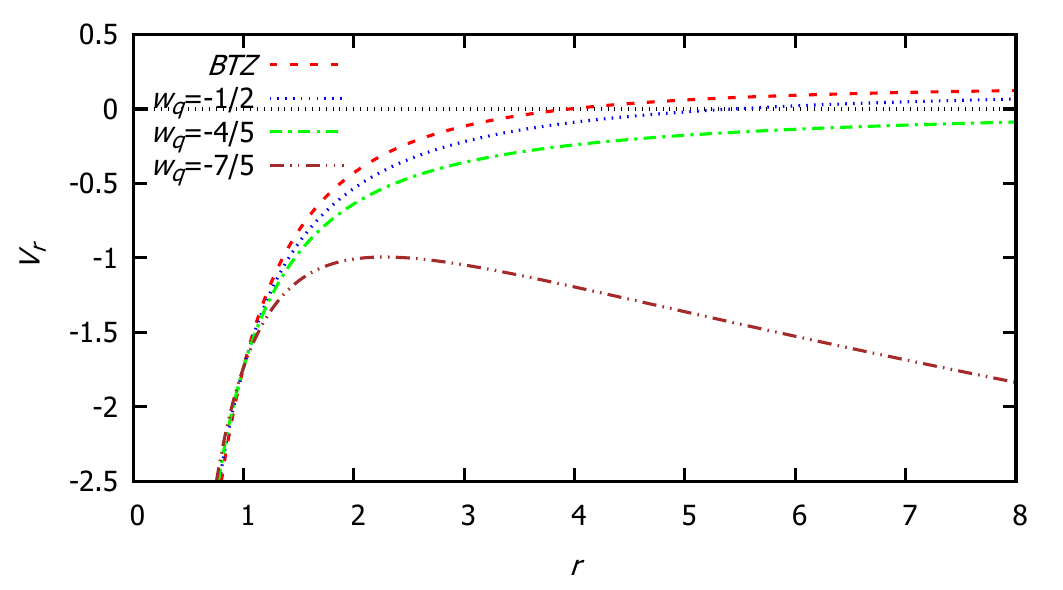}
        \caption{Effective potential as a function of the radial coordinate, $r$, for photons around the 3-dimensional GB black hole. The parameter settings are $\alpha=-0.2$, $q=0.5$, $M=2.0$, $\lambda=1.0$ and $\ell=2.5$, in Planck units, and varying $\omega_q$.}
    \label{Phorb}
\end{figure}
The circular geodesics occur at the points $r_c$ satisfying $\dot{r_c}=0$ and $V'_r (r_c)=0$. Regarding the photon orbits ($\epsilon=0$), we have that
\beq
r_c=\left[-\frac{q (\omega_q+1)}{KM_{eff}}\right]^{\frac{1}{2 \omega_q}}.
\eeq
Notice that such a radius depend on GB parameter, $\alpha$. Moreover, there exist finite circular geodesics for photons only if $q\neq 0$ and $\omega_q<-1$, that is, when the source is phantom-like, since $q>0$. In Fig. \ref{Phorb} we depict the effective potential of massless particles for the GB black hole solutions found above. It is shown that there are circular orbits only for $\omega_q<-1$, which are unstable ($V''(r_c)< 0$).

\section{Thermodynamics}
  The thermodynamic analysis presented here follows the same line of reasoning previously described to obtain the correct expressions for the thermodynamic quantities in the vacuum case. Thus, the entropy remains unchanged and is given by $S = \pi r_h/2$, while the temperature can be easily calculated, yielding
  \beq 
  T_K= \frac{1}{4\pi K}\left(\frac{ 2r_h}{\ell^2}+ 2\omega_qq r_h^{-2 \omega_q-1}\right).
  \eeq 
Thus, the Hawking temperature depends on GB extension of GR. Observing the expression for temperature, we note that it goes to zero for a certain critical value of $r_h$. This behavior suggests the possible formation of a black hole remnant, a stable, non-evaporating object that persists even after the Hawking radiation process has effectively ceased. The size of the remnant is then given by 
\begin{equation}\label{rem}
    r_h^{cri} = (-\omega_q q \ell^2)^{1/2(\omega_q + 1)}.
\end{equation}

\begin{figure}[h]
    \centering
    \includegraphics[width=0.7\linewidth]{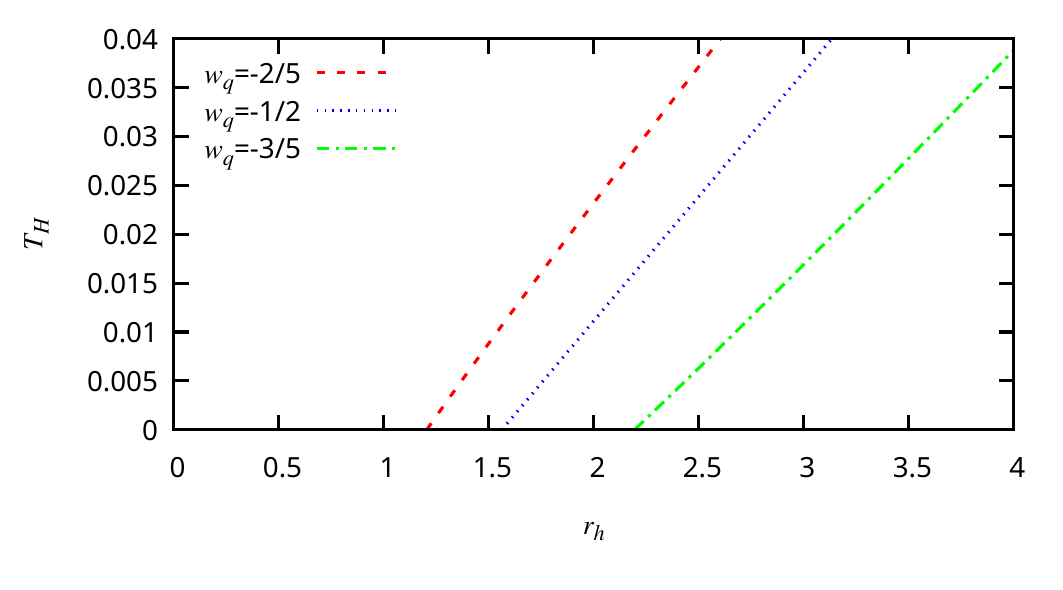}
    \caption{Hawking temperature as a function of $r_h$, fixing $q = 0.5$, $\alpha=-0.2$,$\ell = 2.5$ and varying $\omega_q$.}
    \label{T}
\end{figure}

For $q =0$, it is observed that complete evaporation of the black hole is possible. This suggests that the introduction of quintessence modifies the thermodynamic behavior of the solution, preventing total evaporation and leading to the formation of a stable remnant.
In Fig. \ref{T} we have the plot of temperature as a function of the event horizon radius, where we can see that the critical point at which the temperature goes to zero increases as the value of $\omega_q$ becomes more negative. Therefore, the size of the remnant is directly related to the factor $\omega_q$ of quintessence.

To study the thermodynamic evolution of the black hole, we can calculate the heat capacity, defined as
\begin{equation}
    C = \frac{dQ}{dT_H} = T_H\frac{dS}{dT_H} = T_H\left(\frac{\partial S}{\partial r_h}\right)\left(\frac{\partial T_H}{\partial r_h}\right)^{-1},
\end{equation}
where $dQ$ is the heat emitted or absorbed by the black hole. If the heat capacity is positive, then we have $dT_H <0$, indicating that the temperature decreases and, therefore, the black hole emits heat, i.e., $dQ <0$, and the black hole is stable. On the other hand, if the heat capacity is negative and the temperature decreases, then we have $dQ >0$. In this context, a second-order phase transition is observed when the heat capacity changes sign. Considering that $T_H$ and $\partial S/\partial r_h$ are positive, the sign of the heat capacity will depend exclusively on the sign of $\partial T_H/\partial r_h$. However, since the temperature has no maxima, we conclude that the black hole will not undergo any phase transition. Since $C>0$, we can conclude that the black hole is thermodynamically stable.

Analytically, the heat capacity in terms of the horizon radius is given by
\begin{equation}
   C = \frac{\pi  r_h \left(q \omega_q \ell ^2+r_h^{2\omega_q+2}\right)}{2 r_h^{2 \omega_q+2}-2 q \omega_q (2 \omega_q+1) \ell
   ^2},
\end{equation}
or, in terms of the entropy
\begin{equation}
    C = \frac{q  \pi ^{2 \omega_q+2} \omega_q \ell ^2\, S +4^{\omega_q+1} S^{2 \omega_q+3}}{4^{\omega_q+1} S^{2\omega_q+2}-q \pi ^{2
  \omega_q+2} \omega_q (2 \omega_q+1) \ell ^2}.
\end{equation}

\begin{figure}[ht]
    \centering
    \includegraphics[width=0.65\linewidth]{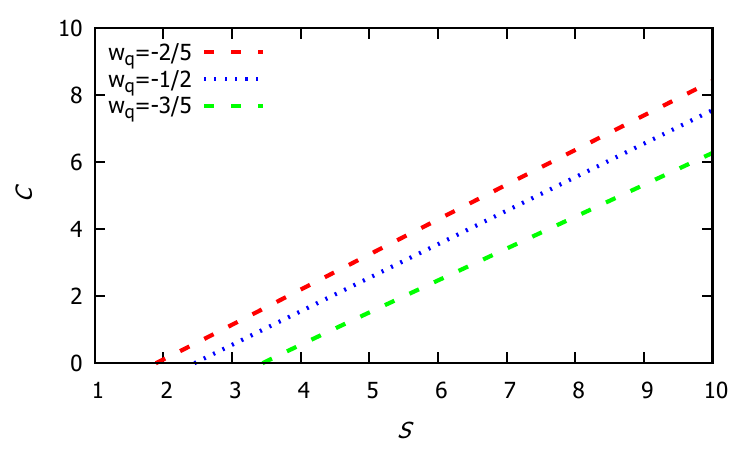}
    \caption{Heat capacity as a function of entropy $S$, for different values of $\omega_q$, fixing $q=0.5$, $\ell=2.5$.}
    \label{fig:heatcapacity}
\end{figure}

The fact that the heat capacity does not explicitly depend on the parameter $K$ can be explained as follows: although $dT_H \propto K$, we also have $dQ = T_H dS \propto K$, so that the contribution of the parameter $\alpha$ eventually cancels out. Graphically, this behavior of the heat capacity can be observed in Fig. \ref{fig:heatcapacity}, where it can be seen that it always takes positive values, considering positive temperatures.

To conclude our thermodynamic analysis, we can also calculate the Helmholtz free energy, defined as \cite{Dehyadegari:2017flm,Dayyani:2017fuz,Wei:2020poh,Junior:2020zdt}
\begin{equation}
    F = M_{eff} - T_HS,
\end{equation}
which gives
\begin{equation}
    F = \frac{1}{4K}\left(\frac{3r_h^2}{ \ell^2} - (4 + \omega_q)qr_h^{-2\omega_q} \right).
\end{equation}

\begin{figure}[h]
    \centering
    \includegraphics[width=0.7\linewidth]{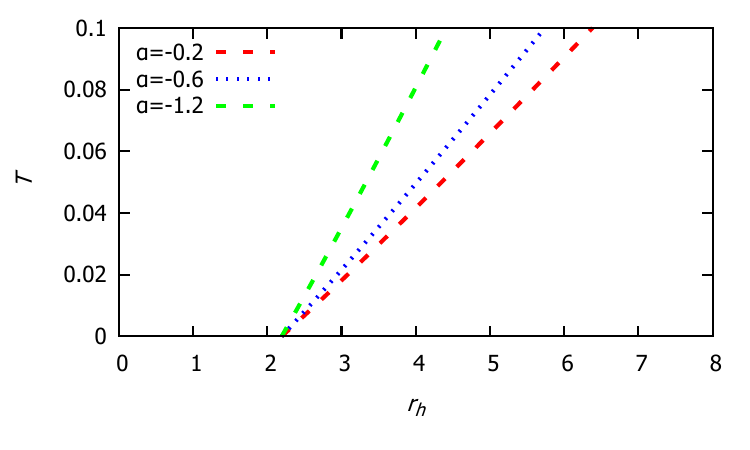}
    \includegraphics[width=0.67\linewidth]{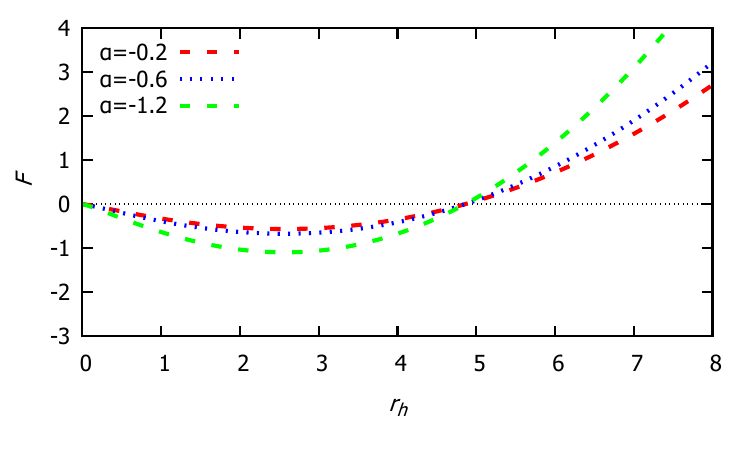}
    \caption{Temperature (top panel) and the free energy (bottom panel) as a function of $r_h$ for different values of $\alpha$, fixing $q = 0.5$, $\omega_q= -3/5$, $\ell = 2.5$.}
    \label{fig:termo}
\end{figure}

With this, we clearly see that the thermodynamic quantities depend on the GB expansion parameter through $K$. In Fig. \ref{fig:termo}, we have plotted the temperature (top panel) and the free energy (bottom panel) as a function of $r_h$ for different values of $\alpha$.

In summary, the black hole requires a remnant whose radius is given by \eqref{rem}. Moreover, since the heat capacity is always positive, this object is locally thermodynamically stable. Additionally, as the free energy changes sign at a certain critical horizon radius where the phase transition (pt) occurs, given by
\begin{equation}
    r_h^{pt}=\left[\frac{\ell^2(4+\omega_q)q}{3}\right]^{\frac{1}{2(\omega_q+1)}},
\end{equation}
we conclude that the black hole, in the final stages of evaporation, is globally stable, undergoing a Hawking-Page–type transition in the process. 
\section{Conclusion}

We have investigated three-dimensional Einstein-GB black holes in the presence of quintessential matter. The derived solutions generalize the BTZ metric while incorporating both higher-curvature corrections and quintessence, reducing to standard cases when either the Gauss-Bonnet parameter $\alpha$ or the quintessence parameter $q$ vanishes. 

Our investigation has yielded several important geometrical and physical insights. The analysis reveals that these solutions possess a single event horizon whose radius exhibits an inverse dependence on the quintessence state parameter $\omega_q$ -- as $\omega_q$ increases, the horizon radius $r_h$ systematically decreases.  A crucial finding concerns the spacetime geometry's behavior at the origin: the presence of quintessential matter ($q \neq 0$) induces a curvature singularity at $r=0$, as evidenced by the divergence in the Ricci scalar. This singular behavior disappears when the quintessence field is absent, restoring regularity at the origin. This singular-to-regular transition demonstrates that quintessential matter fundamentally alters the spacetime geometry near the origin, an effect that persists regardless of the higher-derivative corrections introduced by Einstein-Gauss-Bonnet gravity. 

The study of geodesic motion revealed that stable circular photon orbits exist exclusively for phantom-like quintessence ($\omega_q < -1$), with the orbital radius depends on the parameter $\alpha$. The thermodynamic analysis demonstrated that quintessence significantly modifies the black hole's evaporation process, forming a stable remnant as the Hawking temperature approaches zero. The remnant radius $r_h^{\text{crit}}$ exhibits a monotonic decrease with increasing state parameter values $\omega_q$. This behavior reflects a direct correlation between the black hole's final size and the equation-of-state parameter governing the quintessence field -- as $\omega_q$ approaches the cosmological constant value ($\omega_q \to -1$), the remnant grows in size, while for less exotic matter content ($\omega_q \to 0$), the remnant radius contracts. The black hole remains locally stable from the thermodynamic point of view, as evidenced by the positive heat capacity. 

It is important to emphasize that, contrary to earlier claims, although the Gauss--Bonnet parameter  $\alpha$ does not affect the remnant radius, our results demonstrate that $\alpha$ fundamentally influences all physical quantities, including the horizon structure, geodesics, and thermodynamics, even in the unsourced case. This dependence also persists in the presence of quintessence. In particular, the radius of the circular orbit in geodesic motion exhibits the same sensitivity to $\alpha$.

The fundamental distinction between our solutions of EGB black holes with quintessence in $(2+1)$ and $(3+1)$ dimensions discussed in the literature \cite{Shah:2021rob,Ladghami:2023kyb} lies in the regularization mechanism and the resulting field content: The $D=3$ solution necessitates a scalar field $\phi$ coupled to the geometry to render the Gauss-Bonnet term dynamical, whereas the $D=4$ solutions utilize the Glavan-Lin limit, rescaling the coupling $\alpha \to \alpha/(D-4)$, to retain a pure metric theory. Furthermore, while the quintessence term in the $(2+1)$-dimensional spacetime induces a curvature singularity at the origin and leads to thermodynamically stable remnants with positive specific heat, the $(3+1)$-dimensional counterparts exhibit phase transitions, local instabilities, and logarithmic corrections to the entropy.

These results therefore highlight the rich interplay between higher-curvature gravity and exotic matter fields in (2+1)-dimensional spacetime. The emergence of stable remnants suggests that quintessence may play a crucial role in the final stages of black hole evolution. This work opens new avenues for exploring gravitational theories in lower dimensions with various matter couplings, particularly in the context of black hole thermodynamics and late-stage evaporation dynamics. Finally, regarding the dynamic stability of the solution, we highlight that the analysis of the quasinormal mode (QNM) spectrum is a natural perspective for future work. While our current results ensure thermodynamic stability through the positive specific heat and the formation of stable remnants, a perturbative analysis following standard methodologies (e.g., \cite{Kokkotas:1999bd, Konoplya:2011qq,Lan:2023cvz}) would provide a complementary understanding of the black hole's response to external field perturbations.

\section*{Acknowledgments}
\hspace{0.5cm} The authors would like to thank Conselho Nacional de Desenvolvimento Cient\'{i}fico e Tecnol\'ogico (CNPq) and Funda\c c\~ao Cearense de Apoio ao Desenvolvimento Cient\'ifico e Tecnol\'ogico (FUNCAP) for the financial support.

\bibliographystyle{apsrev4-1}
\bibliography{Ref.bib}
\end{document}